\input{psfig.sty}

\documentclass[useAMS,usegraphicx]{mn2e}
\usepackage{amsmath}
\def\xte{XTE J1814--338} 
\def\ltsima{$\; \buildrel < \over \sim \;$}
\def\simlt{\lower.5ex\hbox{\ltsima}}
\def\gtsima{$\; \buildrel > \over \sim \;$}
\def\simgt{\lower.5ex\hbox{\gtsima}}

\title[Timing of the Accreting Millisecond Pulsar XTE J1814-338]{Timing of the Accreting Millisecond Pulsar XTE J1814-338}

\author[A.Papitto, T.Di Salvo, L.Burderi, M.T.Menna, G.Lavagetto and A.Riggio]{A. Papitto$^{1,2}$\thanks{E-mail: papitto@oa-roma.inaf.it}, T.Di Salvo$^{3}$, L.Burderi$^{4}$, M.T.Menna$^{2}$, G.Lavagetto$^{3}$ and A.Riggio$^{3,4}$\\
$^{1}$Dipartimento di Fisica, Universit\'a degli Studi di Roma ''Tor Vergata'', via della Ricerca Scientifica 1, 00133 Roma, Italy\\
$^{2}$Osservatorio Astronomico di Roma, via Frascati 33, 
Monteporzio Catone, 00040, Italy\\
$^{3}$Dipartimento di Scienze Fisiche ed Astronomiche, 
Universit\`a di Palermo, via Archirafi 36, Palermo, 90123, Italy\\
$^{4}$Dipertimento di Fisica, Universit\'a degli Studi di Cagliari, SP Monserrato-Sestu, KM 0.7, Monserrato, 09042 Italy}
\begin{document}

\maketitle

\label{firstpage}

\begin{abstract} 
We present a precise timing analysis of the accreting millisecond
pulsar {\xte} during its 2003 outburst, observed by {\it RXTE}. A full
orbital solution is given for the first time; Doppler effects induced
by the motion of the source in the binary system were corrected,
leading to a refined estimate of the orbital period,
$P_{orb}=15388.7229(2)$ s, and of the projected semimajor axis, $a\sin
i/c= 390.633(9) $ lt-ms. We could then investigate the spin behaviour
of the accreting compact object during the outburst. We report here a
refined value of the spin frequency ($\nu=314.35610879(1)$ Hz) and the
first estimate of the spin frequency derivative of this source while
accreting ($\dot{\nu}=(-6.7\pm0.7)\times 10^{-14} $ Hz/s). This spin
down behaviour arises when both the fundamental frequency and the
second harmonic are taken into consideration. We discuss this in the
context of the interaction between the disc and the quickly rotating
magnetosphere, at accretion rates sufficiently low to allow a
threading of the accretion disc in regions where the Keplerian
velocity is slower than the magnetosphere velocity. We also present
indications of a jitter of the pulse phases around the mean trend,
which we argue results from movements of the accreting hotspots in
response to variations of the accretion rate. 
\end{abstract}


\section{Introduction}

Millisecond Radio Pulsars have been long believed to be the end
products of long and substantial mass transfer phases on to the neutron
star in a low mass X-ray binary (hereafter LMXB; see e.g. Bhattacharya
\& van den Heuvel 1991). These two classes of objects are linked by
the recycling scenario, that argues how an old and weakly magnetised
neutron star can be effectively spun up to spin periods of few
milliseconds by accretion of matter and angular momentum through a
(Keplerian) accretion disc. Despite the low magnetic fields involved,
these neutron stars are sufficiently fast at the end of the accretion 
phase to switch on again the mechanism that drives the radio pulsar
phenomenon.

The absence of persistent and coherent oscillations in LMXB light
curves represented for a long time an embarrassing problem for the
recycling scenario, until the large collecting capabilities combined
with the unprecedented temporal resolution of the Rossi X-ray Timing
Explorer ({\it RXTE}) satellite allowed Wijnands \& van der Klis
(1998) to discover the first millisecond pulsar in a transient LMXB,
SAX J1808.4--3658 ($P_{spin}\simeq 2.5$ ms, $P_{orb}\simeq2$ h).  

The seven accreting millisecond pulsars that have been discovered so
far (Wijnands 2004; Morgan et al. 2005) have periods ranging from
$\sim 1.5$ ms to $\sim 5.5$ ms, and are all harboured in very compact
binary systems (orbital periods in the $\sim 40$ min to $4$ h range)
with very low mass companions ($\simlt 0.15$ $ M_{\odot}$). Moreover
these objects have always been found in transient systems with more
than $2$ yr recurrence times and generally appear subluminous with
respect to the other LMXBs even during their outbursts, suggesting low
rates of secular mass accretion (Galloway 2006). This feature may
explain why LMXBs generally do not show persistent pulsations if one
lets the accretion rate to control finely the large scale magnetic
field, as this would  be buried under the neutron star surface when
the previously unmagnetised matter accretes too rapidly for the field
to diffuse through it (Cumming, Zweibel \& Bildsten 2001). On the other hand at
lower accretion rates, a magnetosphere can form in the neutron star
surroundings, which channels the transferred matter to the magnetic
poles. The radiation emitted at these spots shows the coherent and
nearly sinusoidal pulsations seen in X-ray light curves of accreting
millisecond pulsars (hereafter AMSP).

The X-ray transient {\xte} was discovered in 2003 during scans of the
central Galactic plane with RXTE (Markwardt \& Swank 2003, hereafter
MS03). This accreting pulsar has a $3.14$ ms spin period and resides
in a binary system, whose orbital period ($4.275$ hr) and minimum
companion mass ($\simeq 0.15$ $ M_{\odot}$) make it the widest and most
massive among all the seven systems discovered so far, that harbour an AMSP.

In this paper we apply timing techniques to the persistent pulsating
activity of {\xte}, in order to investigate the spin frequency
evolution as a result of the balance of the positive torque due to the
accretion of matter and the negative torque due to the interaction
between the magnetosphere and the disc.

\section{Observations and Data Analysis}

\begin{figure}
\includegraphics[width=84mm]{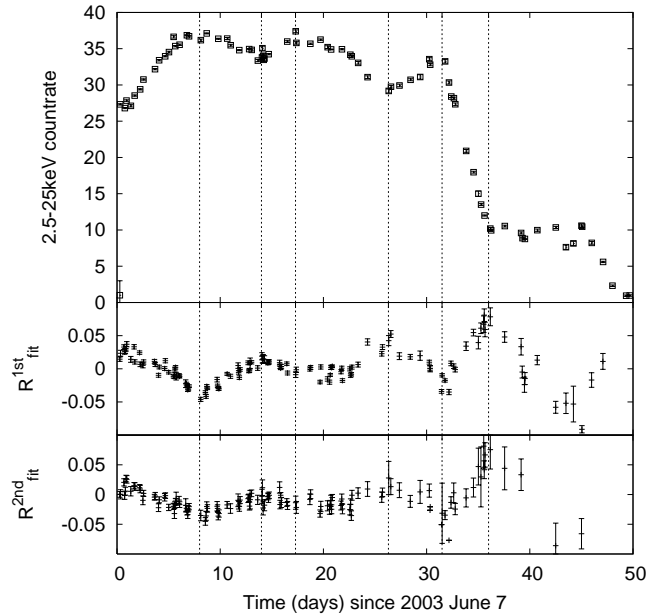}
\caption{ ($2.5$--$25 keV$) light curve of the 2003 outburst of J1814,
as taken by the PCA's Unit 2 aboard RXTE, which was the only one to be
on during the entire outburst. Each point represents the average count
rate in every available observation, which were preliminarily
background subtracted according to the faint source model (top
panel). In the lower panels we plot the residuals with respect to the
 best-fitting constant spin down model of phase delays of the
fundamental ($R^{1st}_{fit}$, middle panel) and of the second harmonic
($R^{2nd}_{fit}$, bottom panel). These plots are
the same as lower panels of Fig.{\ref{timing_fond}} and
Fig.{\ref{timing_harm}} respectively, and are plotted here, together
with dotted vertical lines, to highlight the correlation between the
shape of the light curve and the shape of the residuals of the
fundamental and the second harmonic phase delays, respectively.}

\label{lc}
\end{figure}

{\xte} was found in outburst on 2003 June 5 (MJD 52795); the outburst
lasted for $53$ days, and had a peak (2.5--25 keV) flux of $5\times
10^{10}$ erg/cm$^{2}$/s. After a smooth rise lasting for $\sim 5$
days, the X-ray flux showed three bumps with variations of $\sim 20$ per cent
on a time-scale of $\sim 10$ days,  until it suffered an abrupt cut off
to one fourth of the previous average emission, $33$ days from the
first detection. Afterwards the source fell below the sensitivity
threshold on 2003 July 27 (see upper panel of Fig.\ref{lc} for the
{\it RXTE/PCA} 2.5--25 keV light curve). 

During the coverage of this outburst performed by {\it RXTE}, 28
thermonuclear bursts were observed, exhibiting coherent pulsations at
the same period and strongly phase locked to the persistent pulsations
(Strohmayer et al. 2003). Frequency variability during the course of
the burst has been object of deep study, showing no relevant signs of
departure from the non burst behaviour, similarly to the few Hz
frequency drifts seen in other bursters as SAX J1808.4-3658 (Watts, Strohmayer \& Markwardt 2005). We decided to discard an interval of $200$ s following the
onset of each burst.  We have checked anyway that the inclusion of the
bursts  does not modify significantly the results of the timing.

For the timing analysis we consider data from the RXTE Proportional
Counter Array (PCA, Jahoda et al. 1996), which is made of five
identical units (PCUs) that yield to a total effective area of $\sim
6250$ cm$^2$, sensitive in the 2--60 keV energy band. We used Event mode
data with 64 energy channels and a 125 $\mu$s time resolution. First
of all we corrected the photon arrival times for the motion of the
spacecraft with respect to the solar system barycentre, by using JPL
DE-405 ephemerides along with spacecraft ephemerides. This task was
performed with the {\it faxbary} tool, considering as the best
estimate for the source coordinates the optical counterpart position,
that has a $90$ per cent confidence radius of $0''.2$ (Krauss et al. 2005).  

\begin{figure}
\includegraphics[width=84mm]{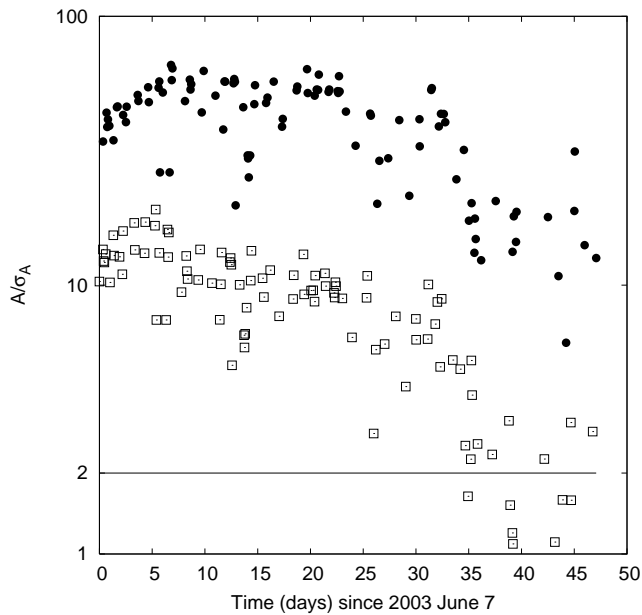}
\caption{Ratios between the fractional amplitudes of the two considered
  harmonic components (filled circles refer to the fundamental
  frequency and squares to the second harmonic) and the relative
  uncertainty $\sigma_A$, plotted against the time since the first
  observation considered. The solid line represents the $2\sigma$
  confidence level.} 
\label{ampli}
\end{figure}

Folding light curves around the spin period $3.18110566967$ ms, we
detected coherent oscillations up to MJD 52844, 47 days after the
first publicly available observation; an harmonic fixed at half of the spin
period is clearly needed to guarantee a good fit of the pulse
profiles. The ratios between the fractional amplitudes, A, of these two
harmonic components and the respective uncertainties $\sigma_A$, are
plotted in Fig.{\ref{ampli}} in order to show a measure of the
significance of each detection. We note that while the fundamental
frequency component is clearly above the 2$\sigma$ limit until MJD
52844, the second harmonic fractional amplitude falls below this
threshold $\sim35$ days after the beginning of the considered
observations, limiting the available number of points to perform the
timing analysis upon this component.

The timing technique is able to reproduce an accurate picture of the
spin behaviour of a neutron star, and is achieved by estimating the
difference between the experimental time of arrival of a given pulse
and the one predicted by using a certain guess of the parameters of
the system, the so-called residual. The evolution in time of these
residuals, depends both on the genuine variations in spin frequency
and on the distance of the guessed parameters from the \textsl{real}
ones (see e.g. Burderi et al. 2006a). Neglecting proper motion and any
relativistic effect, the times of arrival of the photons are affected by
various terms that can be summarised as follows: (a) the motion of the
source with respect to a reference system fixed on the barycentre of
the binary, (b) our inaccuracy in determining the initial spin
frequency, (c) its \textsl{genuine} evolution in time and (d) the
apparent motion of the source induced by the Earth orbital motion,
which arises because of the uncertainty in the source position. In
standard timing techniques all these effects are taken into account
simultaneously, fitting the residuals with a linear multiple
regression of the differential corrections of the relevant parameters
(see e.g. Blandford \& Teukolsky 1976). 

Defining the phase as $\phi=\nu(t-T_0)$, where $\nu$ is the spin
frequency and $T_0$ is the start time of the observation, the
residuals are straightforwardly introduced as the differences between
the predicted and observed phases,
$R_{\phi}(t)=\phi_{pred}(t)-\phi_{obs}(t)$, while their evolution in
time can be expressed with the relation: 
\begin{eqnarray}
\label{eq:res_all}
R_{\phi}(t)=-\phi_0-\delta\nu t - \delta\phi_{\dot{\nu}}(t)+R_{\phi}^{(orb)}+R_\phi^{(pos)}.
\end{eqnarray}
Here $\phi_0$ is a constant, $\delta \nu$ is the correction to the
frequency at the beginning of observations, and
$\delta\phi_{\dot{\nu}}(t)$ is the phase variation induced by a spin
frequency derivative $\dot{\nu}$, that can be expressed as a double
direct integration: 
\begin{eqnarray}
\label{eq:pippo}
\delta\phi_{\dot{\nu}}(t)=\int_{T_0}^{t}\left[\int_{T_0}^{t'} \dot{\nu}(t'')dt''\right]dt',
\end{eqnarray}
so that a constant positive (negative) value of $\dot{\nu}$ leads to a
quadratically decreasing (increasing) term in the phase delays
evolution. The terms induced by uncertainties on orbital parameters,
here labelled as $R_{\phi}^{(orb)}$, and on the source position,
$R_\phi^{(pos)}$, are briefly discussed in the following. 

Burderi et al. (2006a) showed that in the case of millisecond pulsars
the timing approach is greatly simplified by the fact that the
time-scale over which the terms listed above have their effect on
residuals are very different. The orbital period of these systems
($\sim$ few hours) is indeed much shorter than the timescale on which
the spin period derivative seriously affects the phase delays, so that these
two effects can be effectively decoupled. One can thus obtain
corrections to the initial guessed orbital parameters (namely the
projected semimajor axis in light seconds $x=a\:sini/c$ , the orbital
period $P_{orb}$, the time of passage of the NS at the ascending node
at the beginning of the observation $T^*$, and the eccentricity $e$)
simply by fitting with the following relation the modulation that
affects the pulse phases with a periodicity equal to the orbital period
(see e.g. Deeter, Pravdo \& Boynton 1981): 
\begin{eqnarray}
\label{eq:res_orb}
\nonumber
R_{\phi}^{(orb)}&=& \nu x \left[ 
\sin{m} \frac{\delta x}{x} 
- \cos{m}
  \left( m \frac{\delta P_{orb}}{P_{orb}} + \frac{2\pi}{P_{orb}}\delta
    T^*\right) + 
\right. \\ 
&-& \left. \frac{\sin(2m)}{2} \delta e \right],
\end{eqnarray}
where $\nu$ is the spin frequency of the NS,
$m=2\pi(t_{arr}-T^*)/P_{orb}$ is the mean anomaly, and $\delta x$,
$\delta P_{orb}$, $\delta T^*$ and $\delta e$ are the corrections to
the respective parameters. The best orbital solution is found once the
residuals are no longer modulated in this way, rather being normally
distributed around the timing solution with an amplitude
$\sigma_{\phi\:orb}$, that has to be summed in square to the
statistical uncertainty affecting the phase residuals,
$\sigma_{stat}$, to give the overall error on the phase
determination. The expression for $\sigma_{\phi\:orb}$ can be
evaluated simply by differentiating the formula above, substituting the
differentials of the orbital parameters with their errors, and summing
in quadrature: 
\begin{eqnarray}
\nonumber
\sigma_{\phi_{orb}} 
&=&\nu x \left\{
\sin^2m
\left(\frac{\sigma_x}{x}\right)^2
+\cos^2m \left[m^2\left(\frac{\sigma_{P_{orb}}}{P_{orb}}\right)^2 + \right. \right. \\ 
&+& \left. \left. \left(\frac{2\pi\sigma_{T^*}}{P_{orb}}\right)^2 \right]+\sin^2{(m)}\cos^2{(m)}\sigma_{e}^2  \right\}^{1/2}
\end{eqnarray}

We therefore divided each observation in $150$ s time intervals (in
order to avoid a broadening of the folded pulse profiles by the
orbital motion), and then fitted the phase differences according to
the expression (\ref{eq:res_all}), without considering the position
term $R_\phi^{pos}$ (see below), finally obtaining the orbital
solution listed in Table \ref{tab1}. The remaining average uncertainty, that
is $<\sigma_{\phi\:orb}> = 0.016$ ms in our case, fairly matches the
condition $\sigma_{\phi_{orb}}<<\delta\phi_{\dot{\nu}}(t)$, and we can
therefore consider it as a ''timing noise''. The solution we obtain
is more precise than the one already present in literature (MS03; see
Tab.{\ref{tab1}) since in our analysis the entire $47$ days interval
  was taken into consideration, and the accuracy in estimating the
  orbital parameters increases with the length of the time interval
  spanned by the data.

  On the other hand, as the ephemerides of the spacecraft are supposed
  to be known at the highest possible accuracy, the correction for the
  effect induced by the Earth motion on phase residuals depends only
  on the uncertainty that affects the source position. This
  uncertainty induces delays that evolve sinusoidally on $\sim
  P_{\oplus}$ (see e.g. Lyne \& Graham-Smith 1990), so that
  differential corrections on the source coordinates are impossible to
  obtain performing a timing analysis on a such short time baseline
  ($47$ days). Burderi et al. (2006a) also showed how this expression
  can be used to derive an upper limit on the effects of
  uncertainties in source coordinates over the predicted times of
  arrival of X-ray photons, by expanding these delays in series of the
  parameter $\epsilon=2\pi (t-T_0)/P_{\oplus}<<1$. The technique there
  outlined leads to systematic errors affecting the linear and the
  quadratic terms of the time evolution of the residuals, namely the
  spin frequency correction and the spin frequency derivative. In
  particular they found
  $\sigma_{syst\:\Delta\nu}\leq\nu_0y\sigma_{\gamma}(1+\sin^2\beta)^{1/2}(2\pi/P_{\oplus})$
  and
  $\sigma_{syst\:\dot{\nu}}\leq\nu_0y\sigma_{\gamma}(1+\sin^2\beta)^{1/2}(2\pi/P_{\oplus})^2$,
  where $y$ is the semimajor axis of the Earth's orbit in lt-s,
  $\sigma_{\gamma}$ is the radius of the error circle on the position
  of the source and $\beta$ is the ecliptic latitude.

\begin{figure}
\includegraphics[width=84mm]{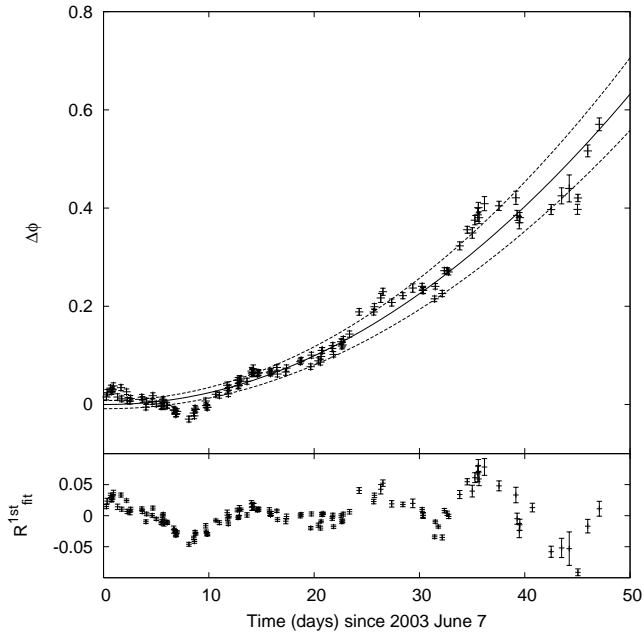}
\caption{Evolution of the pulse phase delays, measured in their natural
  units, computed upon the fundamental frequency component, folding
  every available observation around $P=3.18110566967$ ms. The solid
  line is the best fit constant $\dot{\nu}$ model, while the dashed
  lines represent the contours of the $90$ per cent confidence region  (upper
  panel). Residuals with respect to the best-fitting model
  (lower panel).} 
\label{timing_fond}
\end{figure} 

Having corrected all the arrival times with our best orbital solution,
we could then fold each observation around our best estimate of the
barycentric spin frequency, sampling every pulse profile in 20 phase
bins and finally fitting them with a two harmonics sinusoidal form. The
time evolution of the phase delays, measured in their natural units (time in
units of the folded spin period) of these two components are
showed in Fig.\ref{timing_fond} and Fig.\ref{timing_harm}, exhibiting
a clear and coherent spin down trend, superimposed on which a sort of
modulation is visible. A fit of the fundamental frequency phase delays
evolution with a constant frequency derivative model yields to an
estimate of $<\dot{\nu}_{fund}>=(-6.7\pm0.7)\times 10^{-14}$ Hz/s,
whose uncertainty is quoted at the $90$ per cent confidence level (see below
for details about its derivation). The quality of the fit is greatly
affected by the modulation, whose amplitude can be estimated in $\sim
5<\sigma>_{fund}\simeq0.1$ ms, leading to a very poor quality reduced
chi squared ($=1618/97$). A similar fit performed on the second
harmonic phases delays, this time on a reduced number of points, as some
detections result uncertain 35 days after the first observation
considered here, gives slightly better results ($\chi^2=493/88$),
 as the uncertainties of the measured second harmonic phases are
 on the average larger than the ones of the fundamental, 
while the amplitude of the modulation is just slightly smaller.
The resulting estimate of
the constant frequency derivative in this case is
$<\dot{\nu}_{harm}>=(-8.5\pm0.9)\times 10^{-14} $ Hz/s. The difference
between the obtained values of $\dot{\nu}$ can be attributed to the
different baseline on which the two timing analyses were performed, as
the second harmonic phases are less sampled in the final part of the
outburst. As a matter of fact the restriction to the first $35$ days
of the outburst (an interval along which the two data sets exactly
overlap), yields to values of $\dot{\nu}$ that are comparable
($\dot{\nu}^{0-35}_{fund}=(-8.9\pm0.5)\times 10^{-14}$ Hz/s and
$\dot{\nu}^{0-35}_{harm}=(-8.6\pm0.4)\times 10^{-14}$ Hz/s,
respectively).  

\begin{figure}
\includegraphics[width=84mm]{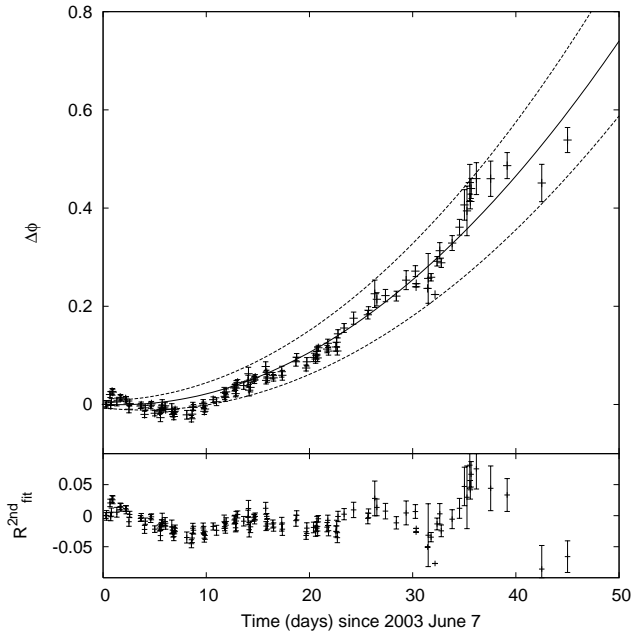}
\caption{ Evolution of the pulse phase delays, measured in their natural
  units, computed upon the second harmonic component, folding every
  available observation around $P=3.18110566967$ ms. The solid line is
  the best fit constant $\dot{\nu}$ model, while the dashed lines
  represent the contours of the $90$ per cent confidence region  (upper
  panel). Residuals with respect to the best-fitting model
  (lower panel).} 
\label{timing_harm}
\end{figure}

Beyond affecting the quality parameter of the least-square fit of the
phase delays evolution, the presence of a modulation superimposed on a
global spin down trend, suggests extreme caution in determining the
uncertainties that affect the determination of the neutron star spin
parameters, which are the frequency at the beginning of the
observation and its mean derivative ($\sigma_{\nu_0}$ and
$\sigma_{<\dot{\nu}>}$ respectively). Rather than considering the
values obtained by a simple least-square fit of the evolution of the
phase delays, taken with their actual uncertainties
($\sigma_{\phi}=(\sigma_{\phi\:stat}^2+\sigma_{\phi\:orb}^2)^{1/2}$),
we amplified all the errors affecting the various points by a common
factor ($3.5$ in the case of the first harmonic) until we obtain
$\chi^2_r=1$ from the fit, and then recomputed $\sigma_{\nu_0}$ and
$\sigma_{<\dot{\nu}>}$ accordingly. As the relative weights of the
single points remain unchanged applying this procedure, the only
result is the amplification of the quoted uncertainties on the
considered parameters. We believe this is appropriate in response to
the presence of residuals that we do not take into account in the
rotational model, but that are not normally distributed around the
timing solution nor seem to be explainable in terms of a wrong
determination of the phases uncertainties (see next section). In this
way we obtain values approximately $3$ times larger for
$\sigma_{<\dot{\nu}>}$, which are the ones we quote throughout this
paper. In Fig.{\ref{timing_fond}} and Fig.{\ref{timing_harm}} the
contours of the $90$ per cent confidence region are plotted with dashed
lines, showing the accuracy with which the global behaviour is
reproduced with these assumptions. Furthermore, considering the $90$ per cent
confidence radius of the position error circle ($0".2$), the apparent
motion of the source affects the determination of $\dot{\nu}$ with a
systematic error $\sigma_{syst\:\dot{\nu}}\leq0.6\times10^{-14}$ Hz/s.

\section{Discussion and Conclusions}

The first aspect we have to discuss is the nature of the phase jiggle
that affects crucially the quality of the fit performed on the phase delays
evolution plots, when either the fundamental frequency or the second
harmonic is taken into consideration. This effect appears as a
modulation around the mean rotational behaviour, as can be easily seen
from the bottom panels of Fig.{\ref{timing_fond}} and
Fig.{\ref{timing_harm}}, which reproduce the residuals in their
natural units from the best-fitting constant spin down model, $R_{fit}$,
in the respective cases. The timescale of this modulation is $\sim12$
days both for the fundamental and the harmonic, ruling out the
possibility that this effect could be yielded by timing errors
referred to a wrong position or orbital solution, as these would make
the phases oscillate with very different timescales ($P_{\oplus}$ and
$P_{orb}$, respectively).

A striking anti-correlation can be found indeed with the shape of the
X-ray light curve of the observed X-ray flux (see Fig.{\ref{lc}}). As a
matter of fact when the flux increases, the residuals $R_{fit}$ of
both the fundamental and the second harmonic decrease, as the source
would be spinning up, and the opposite happens when the flux
decreases, as it can be seen by following dashed
vertical lines in Fig.{\ref{lc}}.  A linear correlation test performed
on the couples of points representing the $2.5-25$ keV count rate and
$R^{1st}_{fit}$ gives a Pearson coefficient $R=-0.80$, which for
$N=113$ points corresponds to a probability of less than $0.01$ per cent for
the points to be uncorrelated. We also performed a rank correlation
test in order to estimate the probability of a monotonic relationship
between the two observables; the Spearman coefficient we obtain is
$\rho=-0.78$ with a similarly low probability of the null
hypothesis. We have also checked that the maximum of the
cross-correlation function between these two time series occurs in
correspondence of non shift of the two series.

An attempt was made to model this behaviour as due to alternating spin
up/spin down states of the NS, directly related to the varying
accretion rate. However, a coherent timing solution could be found
only by allowing the NS to undergo a rather crowded and unlikely
series of discontinuities in the spin frequency. This is because the
instantaneous value of the spin frequency with respect to the folding
one, is represented by the slope of the phase time-evolution curve
($\nu(\overline{t})=d\phi/dt|_{\overline{t}}$), so that a vertex in
the $\Delta\phi$ vs $t$ plot implies a change in the spin frequency,
taking place on a time shorter than the sampling one (see for example
the behaviour of the residuals $R^{1st}_{fit}$ around the day 8 in the
middle panel of Fig.\ref{lc}).

\begin{table}
\centering
\begin{minipage}{84mm}
\caption{Orbital and spin parameters of \xte }
 \label{tab1} 
\begin{tabular}{@{}lll}
\hline
 & MS03 & This work \footnote{Numbers in parentheses are the uncertainties in the last significant figure at the $90$ per cent confidence level.} \\
\hline
$a\sin i/c$ (lt-ms)               & $390.3(3)$   &  $390.633(9)$                   \\ 
$P_{orb}$ (s)                    & $15388.6(3)$ & $15388.7229(2)$     \\
$T^*$ (MJD)                 &   & $52797.8101689(9)$  \\
Eccentricity $e$               &   & $<2.4\times 10^{-5}\;$\footnote{$3\sigma$ upper limit}           \\
& First Harmonic & \\
\hline
$\nu_0$ (Hz) & $314.35610(2)$ & $314.35610879(1)\;$ \footnote{These frequencies refer to the beginning of the observations taken into account, MJD 52797.27387859868. The systematic error driven by position uncertainty is $\sigma_{syst\:\Delta\nu}\leq3\times 10^{-8}Hz$ and was not included in the quoted errors.}
\\
$<\dot{\nu}>$ (Hz/s) & & $ (-6.7\pm0.7)\times 10^{-14}\;$ \footnote{The systematic error driven by position uncertainty is $\sigma_{syst\:\dot{\nu}}\leq6\times10^{-15}Hz/s$ and was not included in the quoted errors.} \\
& Second Harmonic &  \\
\hline
$\nu_0$ (Hz) &  & $314.35610881(1)\;$ $^c$\\
$<\dot{\nu}>$ (Hz/s) & & $ (-8.5\pm0.9)\times 10^{-14}\;$ $^d$\\

\end{tabular}
\end{minipage}
\end{table}

An alternative explanation can be proposed in terms of the motion of
accretion footprints along the NS surface, in response to variations
of the accretion rate. The magnetospheric radius, $R_m$, is commonly
defined as the radius at which the pressure exerted by the magnetic
field equals the ram pressure of the accreting matter, $R_m=\phi
(GM)^{-1/7}\dot{M}^{-2/7}\mu^{4/7}$, where $\mu$ is the magnetic
dipole moment, and $\phi$ is a factor smaller than unity (see
e.g. Gosh \& Lamb 1991; Burderi et al. 1998). This definition gives an
estimate of the radius at which accreting matter leaves the Keplerian
rotational regime and starts to follow the motion of the magnetic
field lines. However, it seems unlikely that this transition happens
sharply, rather taking place through a finite width layer. Along this
transition layer the energy densities of the magnetic field and of the
matter in the disc almost equal, so that we can expect that variations
in the density profile of the disc matter, witnessed by the $\sim10$
days modulation in the observed X-ray flux, become able to bend at
some level the magnetic flux tube along which matter reaches the
hotspots on the NS surface, possibly causing an azimuthal
displacement. As the hotspots rotate with the star this produces the
coherent pulsations that are matter of concern of a timing analysis,
so that a movement of the hotspot position results in lags of the
observed phases. We therefore argue that the jitter in the phase
evolution plots we observe in the 2003 outburst of {\xte} is caused by
motions of the centre of the localised X-ray emission, as a result of
a perturbation of the geometry of the accretion paths related to the
instantaneous accretion rate. This could be particularly the case of a
pulsar that spins down while accreting, if is a close interaction
between the magnetic field lines and the accretion disc that makes the
overall angular momentum flux to be pointed outward (see below). We note that a rather complex behaviour
of the phase delays in response to variations of the instantaneous
accretion rate, is also observed in the case of at least another AMSP,
that is SAX J1808.4--3658, during its 2002 outburst (Burderi et al. 2006b). A
detailed description of the dynamics of the field lines in a regime of
threading of the disc is beyond the scope of this paper, nevertheless
a similar reasoning is proposed by Miller (1996) to explain the phase
lags observed in the Bursting X-ray Pulsar GRO J1744--28 during its
bursts. A similar explanation was also proposed by Galloway et
al. (2001) for the variations of the dip phases in high field pulsars
like GX 1+4 and RX J0812.4--3114. They estimated a
$2^{\circ}-6^{\circ}$ standard deviation of the stochastic wandering
of the accretion column. On the other hand in our case the centre of
the accretion spot would have been varying its longitude with an
amplitude of $\sim15^{\circ}$, and showing strong correlation with the
observed X-ray flux. These differences in the response of the hotspot
longitude to the variations of the accretion rate might be caused by
the higher magnetic fields ($\sim10^{12}G$) of those objects, which
are able to control the flow of matter easier and farther from the
surface. The observation of a similar effect in other accreting pulsars
may provide additional insight on its nature and could
eventually result in an important tool to study the interaction
between the magnetic field and the falling matter, and in particular
the relation between the azimuthal component of the field and the
structure of the accretion disc.

Besides the appearance of this effect of phase jiggle, the general
trend followed by {\xte} is a coherent spin down taking place over the
whole course of the outburst. We consider the value of the spin
frequency derivative computed from the timing on the fundamental
frequency, $<\dot{\nu}>=(-6.7\pm0.7)\times 10^{-14}$ Hz/s, as the most
reliable because it was obtained on the most statistically significant
data set. A spin down behaviour is also displayed by the slowest AMSP
known so far ($P_S=5.4$ ms), XTE J0929--314, whose braking rate during
its 2002 outburst was estimated in $(-9.2\pm0.4)\times10^{-14}$ Hz/s
(Galloway et al. 2002; see also Riggio et al., in preparation, who
report  refined values of the orbital and spin parameters, and in
particular a roughly halved value of the spin frequency
derivative). This spin-down is hard to understand if one lets the
torque released by accreting matter at the magnetosphere to be the
only one acting on the NS, because as far as the disc rotates in the
same sense of the central object the star is expected to be spun up by
accretion. Models developed so far to explain the spin down behaviour
showed by some accreting pulsars mostly point to the braking effect of
the threading of the accretion disc by the magnetic field in regions
where matter in the former rotates slower than the magnetosphere (see
e.g. Gosh \& Lamb 1979; Wang 1987). A similar regime can be attained
in a region close to the inner edge of the disc if, thanks to a low
accretion rate, the magnetosphere is allowed to expand to the
corotation radius, defined as the radius at which the rotational
velocity of the magnetic field lines equals the Keplerian value
($R_{C}=(GM/\Omega_S^2)^{1/3}$). In this case the field lines which
thread the disc beyond the corotation radius spin faster than matter
in its Keplerian motion. Such pulsars are referred to as ''fast'',
because their spin frequency is so high that $R_{C}$ attains values of
the order of the NS radius ($\sim10$ km; $R_{C}=36.3$ km for a
$1.4M_{\odot}$ NS in \xte), leaving a small room for the
magnetospheric radius to remain inside the corotation limit when
responding to a varying accretion rate. In the conventional propeller
picture (Ilarionov \& Sunyaev 1975) accretion is thought to halt
whether $R_m > R_{C}$, because of the building of a centrifugal
barrier that forbids matter to be accreted by the faster spinning
magnetosphere. This scenario has been proven to be not entirely
correct on an theoretical ground, as Spruit \& Tamm (1993) showed how
the velocity excess of the spinning magnetosphere is insufficient to
drive back to infinity all the incoming matter, so that intermediate
disc solutions are expected to arise when $R_m \simgt R_C $. In this
regime matter is not entirely flung out by the centrifugal barrier,
but rather builds up outside $R_m$ and finally accretes on to the
compact object, thanks to the angular momentum losses driven by the
enhanced viscosity, which is assured by the higher densities involved 
and by the
interaction between the magnetic field lines and the accretion
disc. Rappaport, Fregeau \& Spruit (2004) argued how the disc inner
edge around a ''fast'' pulsar could remain fixed at the corotation
radius, even if the nominal value for $R_m$ would place the
magnetospheric boundary beyond it, and showed the existence of
modified Shakura \& Sunyaev thin disc solutions leading to an overall
outward angular momentum flux, with the disc readjusting its structure
to higher densities in its inner parts. Considering a negative torque
of magnetic origin, whose effect is parametrized by a factor $\gamma$,
they obtain an expression for the overall torque acting on an
accreting fast X-ray pulsar:
\begin{equation}
\label{eq:torque}
\tau=\dot{M}(GMR_C)^{1/2}-\gamma\frac{\mu^2}{9R_C^3}
\end{equation}
 where the first term represents the usual spin up torque guaranteed
 by matter accreting at $R_C$, and $\gamma$ is a factor of the order of unity. 

We use the above expression to evaluate the spin down behaviour of
\xte. The bolometric flux was estimated by Galloway, Cumming \&
Chakrabarty (2004), leading to a peak accretion rate of
$6\times10^{-10}$ $M_{\odot}/yr$, where the $\sim8$ kpc distance
estimate was considered (Strohmayer et al. 2003). During the first
$35$ days of the outburst the flux remained loosely constant, so that
at the level of approximation at which eq.(\ref{eq:torque}) was
derived, we could consider its average value on this interval
($5.4\times10^{-10}$ $M_{\odot}/yr$), and use it together with
eq.(\ref{eq:torque}), in order to estimate the magnetic dipole moment
needed to produce such a large spin down. Expressing the overall
torque as $\tau=2\pi I \dot{\nu}$, with $I=10^{45}$ $g\:cm^2$, we
obtain $\gamma^{1/2}\mu\simeq8\times10^{26}$ $G\:cm^3$, that
corresponds to a superficial magnetic field of $B_S\simeq8\times 10^8
\gamma^{-1/2}$ $G$ for a NS of $10$ km of radius. This estimate is
exactly in the plausible range ($10^8-10^9$ G) for the AMSPs to be the
progenitors of radio millisecond pulsars.

\end{document}